\documentclass[a4paper]{article}%,12pt]{report}
\usepackage{graphicx}

\usepackage[T1]{fontenc}
\usepackage[latin9]{inputenc}
\usepackage{color}
\usepackage{natbib} %Harvard style citations Name [Year]
\usepackage{textcomp}
\usepackage{hyperref}%must be the last package included! creates hyperlinks

\begin{document}
%\linenumbers
%Included for Gather Purpose only:
%         input "D:\PhD_Sprites\mybib.bib"

\begin{titlepage}

\begin{center}
\Large Sprite discharges on Venus and Jupiter-like planets: \\a laboratory investigation

\vspace{10 mm}

\large

D. Dubrovin$^1$, S. Nijdam$^2$, E.M. van Veldhuizen$^2$, \\ U. Ebert$^{2,3}$, Y. Yair$^4$, C. Price$^1$\\

\vspace{10 mm}

{\it $^1$ Tel-Aviv University, Tel-Aviv 69978, Israel,\\
\it $^2$ Department of Physics, Eindhoven University of Technology, \\
\it P.O.Box 513, 5600MB Eindhoven, The Netherlands,\\
\it $^3$ Centrum Wiskunde \& Informatica (CWI), P.O.Box 94079, \\
\it 1090GB Amsterdam, The Netherlands,\\
\it $^4$ The Open University of Israel, Ra'anana 43107, Israel.}

\vspace{100 mm}
%\date{\today}
%

 Accepted for publication in\\
 Journal of Geophysical Research - Space Physics,\\
 Special Section related to the Chapman Conference on the \\
 Effects of Thunderstorms and Lightning in the Upper Atmosphere \\
 30 December 2009

\normalsize

\end{center}

\end{titlepage}

\begin{abstract}

Large sprite discharges at high atmospheric altitudes have been found to be physically similar to
small streamer discharges in air at sea level density. Based on this understanding, we investigate
possible sprite discharges on Venus or Jupiter-like planets through laboratory experiments on
streamers in appropriate CO$_2$--N$_2$ and H$_2$--He mixtures. First, the scaling laws are
experimentally confirmed by varying the density of the planetary gasses. Then streamer diameters,
velocities and overall morphology are investigated for sprites on Venus and Jupiter; they are quite
similar to those on earth, but light emissions in the visible range are fainter by two orders of
magnitude. The discharge spectra are measured; they are dominated by the minority species N$_2$ on
Venus, while signatures of both species are found on Jupiter-like planets. The spectrum of a fully
developed spark on Venus is measured. We show that this spectrum is significantly different from
the expected sprite spectrum.

\end{abstract}

%\begin{article}

\section{Introduction}\label{sec: intro}

\subsection{Sprite discharges on earth}

Lightning on earth is often accompanied by electric discharges in the upper atmosphere, known as
TLE's (transient luminous events). The various TLE's observed in the terrestrial atmosphere consist
of several distinct phenomena, which are known as sprites, ELVES, blue jets, as well as several
other sub-species. Red sprites are an impressive display of light above the thunderclouds which
span a vertical range of 50 to 90 km above the surface and take many forms. They are red in color,
although their lowermost, tendril-like part can be blue, e.g. see \cite{Sentman95}.

The mechanism of sprite production on earth is now being understood with increasing precision
(\cite{Pasko97QESheating}, \cite{Raizer98}, \cite{Hu02}, \cite{Hu07}, \cite{Pasko07similarity},
\cite{Nielsen08}, \cite{Luque09}). It is based on the {quasi-electrostatic approximation}. The
charge separation within the thundercloud is slow enough that the conductivity in the mesosphere
and the ionosphere can screen the emerging electric fields. However, a lightning stroke changes the
charge content of the cloud and the surface so rapidly, that the newly generated electric fields
are not screened immediately and appear up to the ionosphere. As the breakdown electric field $E_k$
depends on atmospheric density, it decreases strongly with altitude, and therefore the lightning
generated electric field can exceed the breakdown field at a sufficiently high altitude
(\cite{Wilson25}). This is a necessary condition for the emergence of a sprite a few to several
tens of milliseconds after the parent lightning (\cite{Sao-Sabbas03, Cummer06, Hu07}), but sprites
do not always appear in the mesosphere even if the quasi-electrostatic field is there
(\cite{Luque09}). According to recent triangulations by \cite{Nielsen09}, sprites emerge at
altitudes of 66 to 89~km. High-speed imaging showed that sprites start with downward moving
streamer heads (\cite{Cummer06, Stenbaek07velocity, McHarg07fast}), and telescopic imaging shows
that single channels have diameters of tens to hundreds of meters (\cite{Gerken00morphology}). A
review of sprite properties on earth can be found in \cite{Nielsen08}.

\subsection{Lightning and sprites on other planets}

Lightning discharges are the energetic manifestation of the microphysical and thermodynamical
processes occurring within clouds that reside in a planetary atmosphere. In the solar system,
lightning had been detected by spacecraft via direct optical imaging on Earth and Jupiter, and by
electromagnetic remote sensing on Earth, Venus, Jupiter, Saturn, Neptune and Uranus. Recently,
\cite{Ruf09mars} reported ground-based detection of non-thermal emission from a Martian dust storm,
which they attributed to electrical discharges. No signature of lightning activity had been
discovered on Titan, Pluto and Mercury. We refer the reader to the recent comprehensive reviews on
planetary lightning by \cite{Desch02rev} and \cite{Yair08rev}.
We will elaborate here on the findings that concern lightning on Venus and on Jupiter. On Venus,
lightning activity had been deduced based on the VLF emission detected by the Soviet landers Venera
11 and 12 (\cite{Ksanfomality80}). However, the data from top-side observations by various
spacecraft have not shown un-equivocal optical or electromagnetic signatures, especially after the
fly-byes of the Galileo and Cassini spacecraft (\cite{Gurnett91venus, Gurnett01venus}).
\cite{Krasnopolsky06Venus} reported earth-based measurements of high-resolution spectra of Venus in
the NO band at 5.3~\textmu m and found an NO content of $5.5\pm1.5$~ppb below 60 km altitude. Such
a concentration cannot be explained by cosmic-ray induced chemistry and the suggested mechanism is
production by lightning. \cite{Russell07venus} had analyzed the Venus Express magnetometer data and
inferred a global flash rate on Venus which is comparable to that on Earth, $\sim 50 s^{-1}$. It is
hard to explain how such a high flash rate can occur in the stratiform clouds on Venus. Based on
conventional charge-separation processes which occur in terrestrial thunderclouds,
\cite{Levin83lightning-gen} had shown that the charging rate of the clouds on Venus should be
considerably slower than on Earth, and the resulting flash rate should be of the order of few per
hour. This does not rule out the possibility that other, unknown charging mechanisms do operate
within the clouds, leading to rapid electrification and frequent lightning.
On Jupiter, the Voyager, Galileo, Cassini and New-Horizons missions found clear indications that
lightning discharges are prevalent (\cite{Borucki92jupiter, Baines07jupiter}). They are thought to
occur in the deep H$_2$O clouds that exist in the jovian atmosphere and are estimated to be roughly
100 times more energetic than on Earth (\cite{Yair95}).

Since lightning has been found in planetary atmospheres, it seems reasonable to assume that some
form of TLE, like sprites, occur there as well. \cite{Yair09} estimated the altitude in which
breakdown can occur above the cloud deck in various planetary atmospheres, using the
quasi-electrostatic approximation first proposed by \cite{Wilson25}. \cite{Yair09} predict that for
sufficiently large charge moments, sprites can form in the atmospheres of Venus and Jupiter.

The Japanese Climate Orbiter, Planet-C, planned to be launched in 2010, will search for lightning
on Venus (see \cite{Nakamura07plntC, Takahashi08plntC}). However, the venusian thick cloud layers
might inhibit optical observations. \cite{Yair09} expect that, if lightning exists, sprites may
form above the upper most cloud deck, and thus could be easily observed by an orbiting spacecraft,
both on Venus and on Jupiter. The altitude, diameter, shape and light emission of the observed
sprites could yield valuable information about the charge configuration in the clouds below, as
well as about the gas composition of the upper atmosphere of the planet. In this work we
investigate the expected spectrum of sprites and their morphology via laboratory experiments. This
information can be useful in finding and identifying sprites in upcoming observations.

\subsection{Laboratory experiments on sprites}\label{sec: lab simulation}

\subsubsection{Streamers and Scaling Laws}

It is by now well understood that the large sprite discharges at low air density are essentially
up-scaled versions of small streamer discharges at high air density that dominate the initial
breakdown of large gas volumes in a sufficiently strong electric field. As discussed by
\cite{Ebert06review, Pasko07similarity, Briels08sim, Ebert09} and supported by observations by
\cite{Gerken03morphology, Marshall06scale, McHarg07fast} and by simulations by
\cite{Liu04photo-ion, Pasko06theormodel, Luque07, Luque08interaction, Luque09}, streamers and
sprites are essentially related through similarity or scaling laws. Similarity laws for discharges
in gasses of the same composition, but of different density were probably first formulated by
Townsend for the so-called Townsend discharge at the beginning of the 20th century; they are
discussed in many textbooks of gas discharge physics. While there are also many deviations from
similarity in other discharges, similarity in the propagating heads of streamer discharges holds
particularly well because these fast processes are dominated by collisions of single electrons with
neutral molecules, while two-step processes and three-particle processes that would be density
dependent, are negligible on these short time scales. This implies that the basic length scale of
the streamer discharge is the mean free path of the electron $\ell_{mfp}$ which is inversely
proportional to the density~$n$ of the medium, $\ell_{mfp}\propto1/n$. Similarity at varying
density $n$ implies that the streamer velocity, as well as velocity and energy distributions of
individual electrons are the same at similar places, while length scales as $\ell \propto 1/n$,
electric fields as $E\propto n$ etc. The similarity of the overall morphology including diameters
and velocities of streamers and sprites in terrestrial air of varying density were recently
confirmed experimentally by \cite{Briels08sim} (see also discussion by \cite{Ebert09}).

The basic arguments for the similarity laws between streamers at different gas densities do not
depend on the gas composition, and indeed the similarity laws were recently confirmed
experimentally by \cite{Nijdam10pure-gasses} for widely varying mixing ratios of N$_2$ and O$_2$
and for argon. This opens a possibility to simulate planetary sprites through laboratory
experiments on the corresponding gas mixtures. To the best of our knowledge such simulations were
never performed in gas mixtures specifically chosen to simulate planetary atmospheres other than
earth. We here investigate sprites on Venus and Jupiter by means of creating streamer discharges in
gas mixtures that correspond to the planetary atmospheres of Venus and Jupiter respectively,
CO$_2$-N$_2$ (96.5\%-3.5\%) and H$_2$-He (89.8\%-10.2\%), respectively. The mixtures' compositions
are based on the NNSDC (compiled by Williams,
\url{http://nssdc.gsfc.nasa.gov/planetary/planetfact.html}).

It is well known that a minimal voltage is required to start streamer discharges from a needle
electrode. This voltage is called the inception voltage; it depends on electrode shape and material
as well as on gas composition and density and (up to now) has no direct interpretation in terms of
microscopic discharge properties. The streamers that are formed at the inception voltage have a
minimal diameter, but if the voltage rises rapidly to a voltage higher than the inception voltage,
considerably thicker and faster streamers can emerge, which eventually break up into more and
thinner streamers until the thinnest streamers again have the minimal diameter
(\cite{Briels06circuit, Briels08prmtrs}). Therefore sprite tendrils should also attain a minimal
diameter, even though a sprite does not, of course, emerge from a needle electrode
(\cite{Luque09}). The minimal diameter at each pressure is a convenient quantity to test the
similarity laws as the product of minimal diameter and density (\textit{reduced minimal diameter})
should not depend on density (\cite{Briels08sim}). In section~\ref{sec: d and v}, we confirm the
similarity laws for positive streamers in gas mixtures that correspond to the atmospheres of Venus
and Jupiter.

\subsubsection{Sprite spectrum}

In looking for planetary sprites, the expected spectrum of their optical emissions is of major
importance. Such knowledge allows identifying sprites and constructing observation devices.
Laboratory settings are convenient for performing such measurements. Terrestrial sprites' spectrum
was successfully simulated by \cite{Williams06radiance}, who have created a ``sprite in the
bottle'' in a glow discharge tube. This experiment was repeated by \cite{Goto07sprites}.  The
spectrum measured by this method agrees with the spectrum of terrestrial sprites, measured by
\cite{Mende95Spectrum} and \cite{Hampton96spectrum}. To our knowledge, the simulation of planetary
sprites' spectra was not attempted yet.

Sprites are transient discharges, believed to be equivalent to streamers at higher pressures,
therefore a laboratory streamer discharge should simulate a sprite much better than a glow
discharge. Both streamer and glow are so-called low temperature discharges (i.e., the temperature
of the neutral gas molecules does not increase much within the plasma). A streamer is a very
transient process, where an ionization front moves rapidly within a locally enhanced electric field
that well exceeds the breakdown value. At the same time the electric field outside the streamer
head can be significantly weaker than the breakdown field. While creating an ionized trail behind
it, only the moving streamer head emits light. The glow discharge, on the other hand, is
stationary; therefore excitation levels, chemical reactants and ions that can be generated in
several steps, are present. The entire glow column is ionized and emits light, and the electric
field within it is well below the breakdown field.

In our experiments, the discharges are created by voltage pulses of $\mu$s duration. The streamers
emit more light at lower pressures, where they also easily transit into stationary glow, that is
even brighter. These are the conditions where the spectrum can be measured in a reasonable time
span. Thus the spectrum we report in section~\ref{sec: spectrum} is in fact the spectrum of a
discharge that starts as a streamer and quickly transforms into a glow column. Recent experiments
with a new electric circuit able to generate much shorter voltage pulses that create streamers that
do not turn into glow, have shown that the spectra of streamer discharges and of short pulsed glow
discharges are quite similar in pure nitrogen at pressures between 25 and 200~mbar (Nijdam
\textit{et al.}, \url{http://arxiv.org/abs/0912.0894v1}, version 1). To suppress multi-step
processes that do not occur in a streamer head, we use a low frequency repetitive discharge, with a
frequency of 1 or 10~Hz, rather than a stationary or high frequency discharge. In a low frequency
repetitive discharge, such as the one described here, there is enough time for the gas to return to
its original neutral state, or for the reaction products to be flushed away by the gas flow, if
present.

We remark that several research groups have measured spectra of hot plasmas to simulate lightning
in planetary atmospheres. \cite{Borucki96spec} created hot plasma using a laser pulse and measured
the emission spectrum. The conditions in that experiment resemble those in the hot plasma lightning
channel, where many chemical processes take place due to high temperature rather than due to a high
electric field. Goto and Ohba (unpublished report, 2008), simulated the lightning spectrum in pure
CO$_2$ using spark discharges. In section~\ref{sec: spark} we report our measurements of the
spectrum of sparks in the mixture that represents Venus' atmosphere. The spectrum of the cold
plasma streamer and glow discharge is significantly different from that of the hot plasma spark
discharge in this mixture, and the same is probably true in other gas mixtures. Unfortunately we
have not been able to produce a spark in the jovian mixture with our set-up. We observed that the
glow discharge became stronger, but did not transform into a spark discharge. Due to time
limitations, we did not investigate the reason for this.

\subsection{Outline}

In the following we review our experiments on two planetary gas mixtures, CO$_2$-N$_2$ representing
Venus and H$_2$-He representing Jupiter. In section~\ref{subsec: set-up} we give a description of
the experimental setup, and in section~\ref{subsec: overview} we give a qualitative description of
the discharges observed. We discuss the qualitative similarities and differences between the two
planetary gasses and dry air. Dry air was examined in previous experiments by \cite{Briels08sim}
and \cite{Nijdam10pure-gasses}. Section~\ref{sec: d and v} deals with scaling laws in the two
planetary gasses. We measure the {\it reduced minimal diameter} of streamers and their velocity. We
are able to confirm that scaling laws apply to our gasses, based on the reduced diameter
measurements. In section~\ref{sec: spectrum} we present the spectra we measured in streamer and
glow discharges in the two gasses at a pressure of 25~mbar, and the spark spectrum in the Venus
mixture, that was measured at a pressure of 200~mbar.

\section{Experimental set-up and overview of the discharges}\label{sec: set-up}

\subsection{Set-up}\label{subsec: set-up}

\begin{figure}
    \includegraphics[width=9cm]{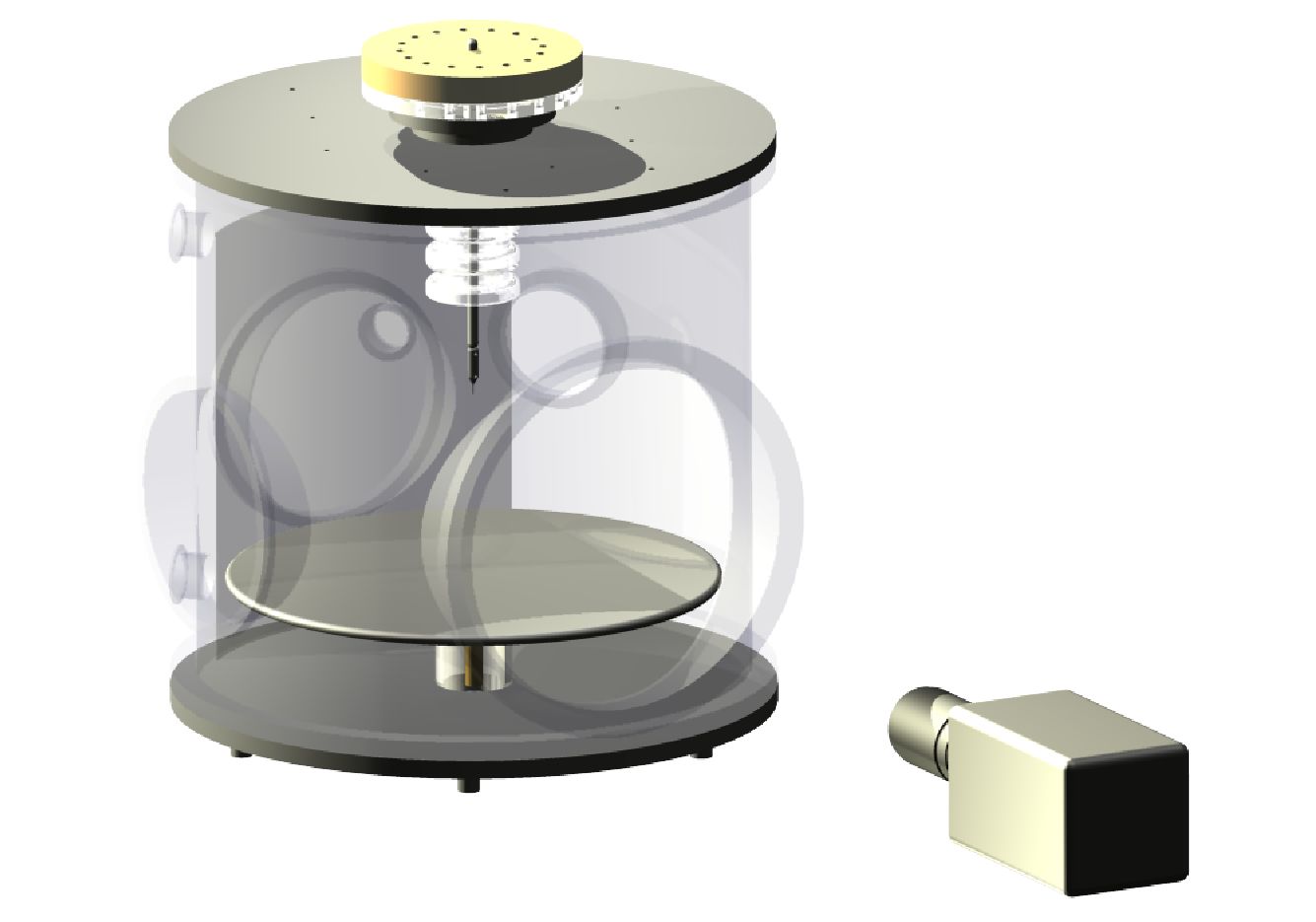}
    \caption{
      Overview of the high purity vacuum vessel with the ICCD
      camera. The wall of the vessel has been rendered transparent in the figure so that the anode tip
      and cathode plane are clearly visible.}
      \label{fig:Setup-overview}
\end{figure}

Positive streamers are created in a large cylindrical stainless steel vacuum vessel with an
internal diameter of 50 cm and an internal height of 30 cm. The vacuum vessel contains a sharp
tungsten tip, placed 16~cm above a grounded plate. The whole vessel is placed inside a Faraday
cage. A large quartz window is positioned on the vacuum vessel. The Faraday cage contains a window,
covered by a conducting layer of Indium-Tin-Oxide (ITO), through which the discharges can be
photographed by a camera outside the Faraday cage.  This window is transparent to visible light,
but not to UV radiation of wavelengths below 300 nm. The streamer discharge is imaged by a Stanford
Computer Optics 4QuickE ICCD camera through a Nikkor UV 105~mm f/4.5 lens. The imaging set-up is
sensitive to wavelengths from 300 to 800 nm. A schematic drawing of the vacuum vessel with the
camera is given in Figure~\ref{fig:Setup-overview}.

We have used a set-up that is specifically designed to ensure the purity of the gasses inside. For
this reason, the set-up can be baked to reduce out-gassing, it contains no plastic parts, except
for the o-ring seals and it stays closed all the time. When not in use, the set-up is pumped down
to a pressure of about~$2\cdot10^{-7}$~mbar.

During the experiments, we have used constant pressures between 25 and 1000~mbar. The gas inside
the set-up is flushed constantly. The absolute flow rate is controlled by a mass-flow controller
and depends on pressure. The flow rate is chosen so that all the gas is replaced every 25~minutes.
This ensures that the contamination caused by out-gassing or leaks is significantly below 1~ppm for
all pressures used. We use two different gas mixtures that are pre-mixed by the supplier: a mixture
(Venus) that consists of 96.5\%~CO$_{2}$ and 3.5\%~N$_{2}$ and a mixture (Jupiter) that consists of
89.8\%~H$_{2}$ and 10.2\%~He. According to the specifications, the contamination level is below
1~ppm for both mixtures. As an extra safety measure, we never fill the vessel with more than 800
mbar of the flammable H$_2$-He mixture.

\begin{figure}
    \includegraphics[width=9cm]{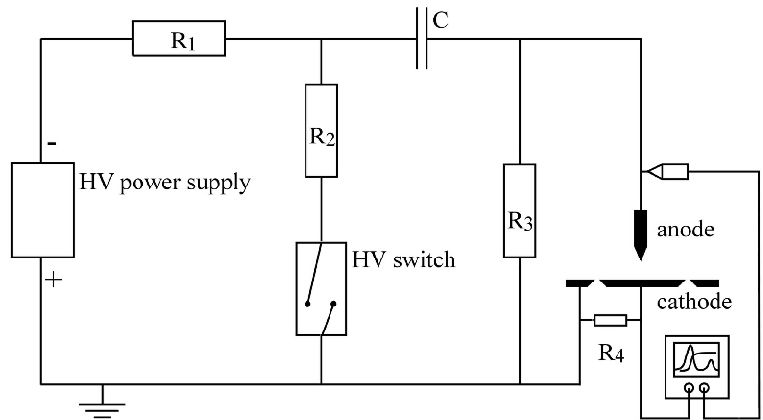}
    \caption{Schematic depiction of the electrical circuit. The HV switch is a spark gap. The figure is
             taken from \cite{Briels06circuit}.}
    \label{fig:Circuit-simple}
\end{figure}

During a measurement, a voltage pulse is applied to the anode tip. A 1~nF capacitor is charged by a
high voltage negative DC source. Now a trigger circuit triggers a spark gap, which acts as a fast
switch. The capacitor is discharged and applies a positive voltage pulse on the anode tip. Positive
streamers are initiated near the anode tip and propagate in the direction of the plate. This pulse
has a rise-time of about 15~ns and a fall time of about 10~\textmu{}s, depending on the choices for
the resistors and the streamer discharge itself. See Figure~\ref{fig:Circuit-simple} for a
simplified drawing of the circuit. In most cases, a repetition rate of 1~Hz is used.

In the images presented in this work, the original brightness is indicated by the multiplication
factor MF as in \cite{Nijdam10pure-gasses}. This value is a measure of the gain of the complete
system, it includes lens aperture, ICCD gain voltage and maximum pixel count used in the
false-colour images. An image with a high MF value, is in reality much dimmer than an image with
similar colouring, but with a lower MF value. We have normalized the MF values in such a way that
the brightest image presented here has an MF value of~1.

More information about the circuit, discharge vessel, experiment timing, imaging system and
measurement techniques can be found in \cite{Briels08prmtrs, Briels08sim, Nijdam2009}.

\begin{figure}
    \includegraphics[width=12cm]{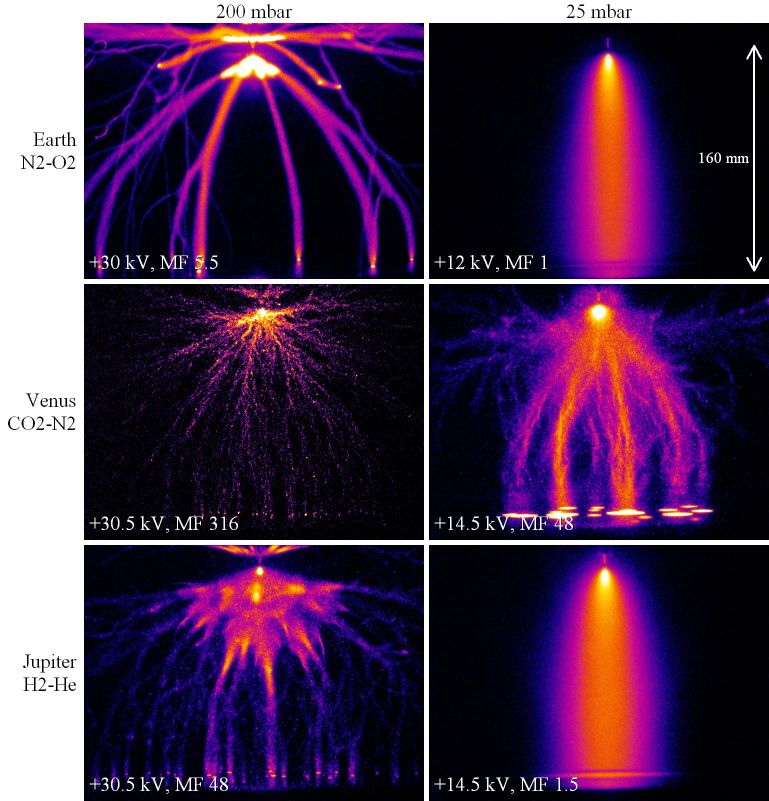}
    \caption{
        Comparison of discharges in air, and in the two gas mixtures. The pressures are the same in each
        column. The applied voltage is indicated in each figure.
        The color scheme of the raw images is modified to enhance the filamentary structure of the
        streamers.
        This is indicated by the multiplication factor (MF).
        These images were taken with long exposure times,
        such that the full development of the discharge was recorded,
        including late streamers and glow.
        In the similarity measurements we focus on the primary streamers, and use much shorter exposure times.
        Images in air were taken by F.~M.~G.~H.~van~de~Wetering (unpublished report
        2008).
    }
  \label{FIG: overview}
\end{figure}

\begin{figure}
  \includegraphics[width=12cm]{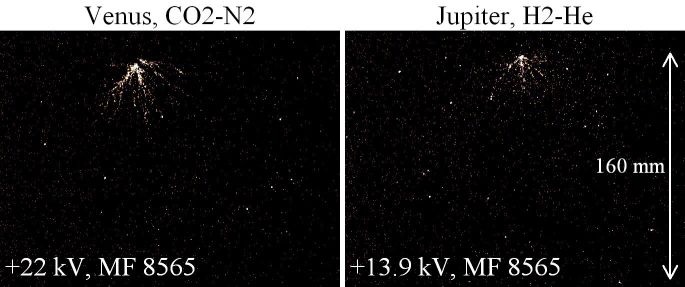}
  \caption{
    Streamers at inception voltage and a pressure of 800 mbar.
    The applied voltage and the multiplication factor are indicated in each image.
    The exposure time is longer than the time required for the full development of the streamer.
    The full development of the streamer is recorded in both images.
    The indicated applied voltages are a few~kV higher than the inception voltage.
    The light emitted by the streamer is very weak, which required the use of the maximum gain
    voltage (950V) in our camera. This accounts for the many specs in the images, which are
    most likely due to noise in the ICCD camera.
    }
  \label{FIG: 800 mbar}
\end{figure}

\begin{figure}
  \includegraphics[width=12cm]{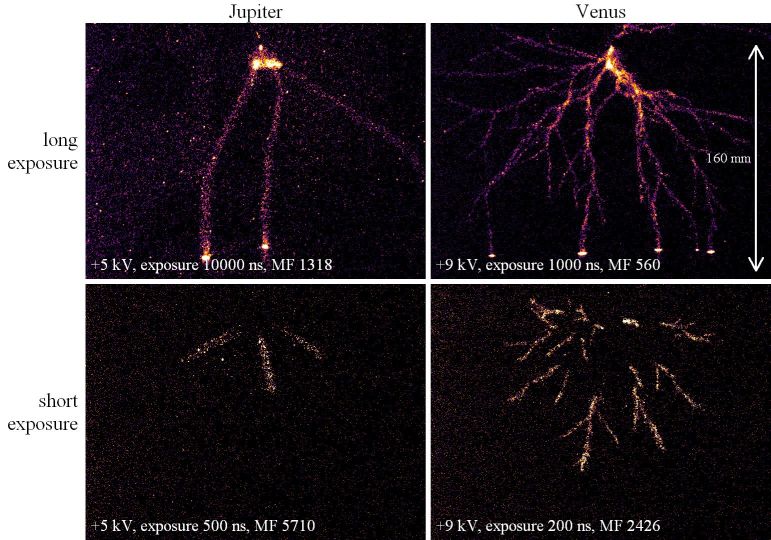}
  \caption{
    Streamer images at minimal inception voltage and a pressure of 50 mbar.
    The applied voltage, the exposure time and the multiplication factor are indicated in each image.
    The top row shows the entire streamer development.
    Under the specified conditions very little current flows through the ionized channels.
    The bright dots at the bottom of the picture are the points where the streamers reach the plate electrode.
    Most of the streamers propagate in the direction of the plate, but some of them propagate to the vessel's walls.
    The bright area at the top of the images is the needle electrode and the glow around it.
    The bottom row shows snapshots of the streamer development, taken with a delay of a few hundred
    nano-seconds after the initiation of the streamer.
    Note the  shorter exposure times, relative to the top row.
    Such images are used in the calculation of the minimal diameter and the velocity,
    as described in section~\ref{sec: d and v}.
    }
  \label{FIG: strmrs}
\end{figure}

\subsection{Overall structure of the discharge}\label{subsec: overview}

We were able to observe streamers in both gas mixtures. We show examples of the  streamers we
observed in Figure~\ref{FIG: overview}. The voltages in this figure are well above the inception
voltage, the minimal applied voltage that is required to create a streamer. Examples of streamers
close to the inception voltages are shown in Figures~\ref{FIG: 800 mbar} and~\ref{FIG: strmrs}. The
images clearly illustrate the fact that a higher voltage is required to create a discharge when the
pressure is increased. Increasing voltage, or decreasing pressure causes the discharge to be
brighter and less filamentary. When planetary mixtures are compared, we notice that the initiation
of streamers in both our gasses requires higher voltages than in dry air at the same pressure. In
the venusian mixture streamers are particularly difficult to create. For example, at~800 mbar a
minimal voltage of approximately 20~kV is required to initiate a streamer discharge. At this
voltage the streamer dies out before it bridges the gap. At the same pressure, streamers in the
jovian mixture are created at lower voltages below$\sim$15~kV. These streamers do not cross the gap
either.  Likewise, streamers at 1~bar in dry air can form when the applied voltage is as low as
$\sim$10~kV, as was shown by \cite{Briels08prmtrs} in a similar set-up.

Only the streamer's head produces light. Therefore an image taken with a long exposure shows the
trace left by the streamer head as it passes. With a shorter exposure just a section of the
streamer's path is seen (see Figure~\ref{FIG: strmrs}). At lower pressures, or higher voltages, the
streamers develop into glow after reaching the plate electrode, resulting in a more or less uniform
light that is emitted from the entire channel. This is seen in the right column of Figure~\ref{FIG:
overview}. The glow phase lasts longer than the streamer phase, and emits significantly more light.
Both in dry-air and in the jovian mixture, H$_2$-He, a single wide channel is formed when the
pressure is below 50~mbar. In the venusian mixture, CO$_2$-N$_2$, the filamentary streamer
structure persists at the lowest pressures. Several channels conduct most of the current in the
glow phase, as indicated by their strong intensity in the images. The light emission is usually too
weak to be observed by the naked eye, except at the lowest pressures, below 100~mbar. In the
venusian mixture the streamers tend to be bluish-green in color, and in the jovian mixture they
seem to be pink.

The images in Figure~\ref{FIG: overview} are long exposure images taken at voltages that are
relatively high when compared to the minimal inception voltages at the corresponding pressures.
Under such conditions the streamers bridge the gap and conduct enough current to create the glow
phase. When voltage is set as close as possible to the inception voltage value, the glow phase does
not appear. An example is shown in Figure~\ref{FIG: strmrs}, where the top row is a full exposure
image, that shows the streamer channels, and no indication of a glow phase. Note the different
applied voltages, indicated at the bottom of each image, in the venusian the inception voltage is
higher than in the jovian mixture. This difference persists in other pressures as well. We note
that the jovian streamers have much less branches than the venusian streamers, under very similar
conditions. Together with the high inception voltages required, this may be an indication of lower
ionization rates in the venusian gas mixture. Apparently, streamer heads are less stable in this
mixture, which causes them to branch more often. In the top right image of Figure~\ref{FIG: strmrs}
we observe many branches that stop in the middle of the gap, never reaching the other side. There
are much less such branches in air and in the H$_2$-He mixture. Looking closely at the streamer
channels one sees that in the venusian mixture they are much more jagged than in the other gasses,
and are almost never straight. More images of streamers in air are available in e.g
\cite{Briels08prmtrs} and \cite{Nijdam10pure-gasses}.

The morphological differences between the streamers in the two mixtures are probably due to two
properties of the gasses, $(i)$ their atomic or molecular structure, and $(ii)$ the photoionization
mechanism. We here recall the basic mechanisms, that will have to be elaborated further in future
work. $(i)$ A noble gas like He or Ar consists of single atoms, N$_2$ and H$_2$ are molecular
gasses consisting of two atoms, and CO$_2$ consists of three atoms. The molecular gasses, and in
particular CO$_2$, have many rotational and vibrational states at low energies that can absorb the
energy of colliding electrons; therefore the electrons experience more friction than in noble
gasses. Furthermore, He and H$_2$ have only two electrons each which results in only few electronic
excitation states. For this reason, streamer propagation is much easier in jovian than in venusian
or terrestrial atmospheres.  $(ii)$ Positive streamers as investigated here move approximately with
the electron drift velocity, but against the direction of electron drift; therefore they depend on
a source of free electrons ahead of the ionization front. These free electrons can be supplied
either by a nonlocal photoionization effect or by background ionization. Researchers currently
generally agree that photoionization is the dominant effect in nitrogen-oxygen mixtures like
terrestrial air (see~\cite{Zheleznyak82photo-ion, Liu04photo-ion, Pancheshnyi05photo-ion,
Ebert06review} and references therein), though \cite{Pancheshnyi05photo-ion} argues that high
repetition frequencies can shift the balance towards background ionization as well. However, in any
other gas the propagation mechanism of positive streamers is not really understood. In pure oxygen
or nitrogen or argon, it is presently under investigation experimentally by
\cite{Nijdam10pure-gasses} and theoretically by Wormeester (manuscript in preparation). These
investigations suggest that photoionization is able to stabilize wide streamer heads while positive
streamers propagating due to background ionization are less straight and branch easier. These
considerations may influence both the streamer head velocity and diameter, as well as other
factors, as shown by the results described in section~\ref{sec: d and v}. Our preliminary
investigations on the photoionization mechanism in venusian or jovian atmospheres below support
this scenario.

\cite{Teich67photo-ion} and later \cite{Zheleznyak82photo-ion} attribute the photoionization
process in nitrogen-oxygen mixtures to several nitrogen emission lines in the wavelength range
98-102.5~nm. These photons are energetic enough to ionize oxygen molecules at~$\sim$12~eV. This
photoionization takes place after some travel distance that depends on the oxygen concentration;
thus it is a nonlocal process. On Venus, CO$_2$ requires a higher ionization energy of~$\sim$13~eV,
however the nitrogen molecule does not have emission lines that are energetic enough to ionize this
molecule in a one-step process. Therefore other, less effective processes must account for the
streamer propagation in the venusian mixture. With the low photoionization efficiency in the
streamer head, it becomes less stable and more likely to branch.
The picture in the jovian H$_2$-He mixture is completely different. There is a large gap between
the ionization energies of the hydrogen molecule and helium. A photon with a wavelength below
77.5~nm is energetic enough to ionize the hydrogen molecule at~$\sim$16~eV. The helium atomic
spectrum has a group of lines in the range 50.7-58.4~nm that can produce photons with enough energy
to ionize the hydrogen molecule in a one step process. Assuming that the photoionization process is
effective in the H$_2$-He mixture, one is not surprised to find that the streamer heads are as
stable as they are. Ionization and atomic spectral data in this paragraph are taken from the NIST
database (\cite{NIST-ASD} and \cite{NIST-chem-book}).

The optical brightness of the streamers in both gasses is considerably weaker than in the gasses
used in previous experiments (ambient air, nitrogen--oxygen mixtures and argon). Our setup is not
calibrated to estimate absolute optical brightness, however a rough estimation of the relative
intensity of planetary streamers to streamers in air is possible. The pixel gray level in an image
is proportional to the light intensity and depends on the camera settings We can estimate the
ratios between average intensities per area in different gasses, with the following method: we
{record short} exposure images using the same equipment and under the same pressure and voltage
conditions. We evaluate the averaged gray level per area of the streamer sections in these images.
Our estimation shows that the optical brightness of both Jovian and Venusian streamers appears to
be about a 100 times weaker than of streamers in air at similar pressures. A similar method is used
to estimate the brightness of terrestrial sprites \cite{Yaniv09} Averaged intensity per area
depends on pressure, at lower pressures streamer images are brighter in all mixtures. These
findings depend on equipment choice, since they are measured with the specific wavelength
sensitivity curve of our equipment. So a very bright line at the edges or outside of this curve is
not observed, but could be observed when other equipment is used.

\section{Measuring Diameter And Velocity}\label{sec: d and v}

\subsection{Method}\label{subsec: d and v method}

We test the similarity laws discussed in section~\ref{sec: lab simulation} by determining the
diameter and the velocity of the minimal streamers in planetary mixtures. In search for the minimal
streamer diameter and velocity, we use images taken as close as possible to the inception voltage,
the minimal voltage that is required to create a discharge (see \cite{Briels06circuit} for
details). We determine this minimal voltage by gradually sweeping up the applied voltage, until
streamers begin to appear in the discharge gap. At first streamers appear sporadically once every
few pulses. These discharges are very difficult to image. For this reason we often apply a voltage
that is slightly higher than the minimal inception voltage, where the discharge appears quite
regularly. We note that The discharges are not absolutely independent. If a pulse triggers a
discharge, the chances that the next pulse will trigger a discharge as well are increased. This
suggests that some residue ionization is left in the gas for at least one second after the
discharge. The discharge itself lasts several hundreds nano-seconds.

The streamer diameter is determined from the recorded images with the method described by
\cite{Briels08prmtrs}: One to five of the thinnest streamers are chosen in each image. In choosing
the streamer sections we keep in mind that not all the streamers in a given image are in focus.
Those streamers that are not, will appear to be wider than those that are in focus, \textit{and
will not be selected for diameter determination}. Several perpendicular cross sections of each of
the chosen streamers are taken. These cross sections are averaged so that they form one single
cross section per streamer. The diameter of the streamer is determined as the full width at half
maximum (FWHM) of the averaged cross section. The diameters of all the chosen streamers are then
averaged. We remark that the streamer images in the planetary gasses have a relatively low signal
to noise ratio. This makes the determination of FWHM of a single cross section almost impossible.
Averaging over many cross sections, and the extraction of the FWHM of the averaged line, gives
quite good results.

We use the following method to measure velocity, as described by \cite{Briels08prmtrs} :We take
short exposure images of streamers while they propagate in the middle of the gap, far away from
both electrodes, where the effect of the electrodes on the field is minimal. The image shows the
path that the streamer heads have crossed within the exposure time, hence their velocity can be
determined by dividing the length of the streamer by the exposure time. We choose the longest
straight streamer sections in each image, which do not branch. These streamers are most likely to
have propagated more or less in the image plane, with approximately constant velocity.

\subsection{Results}

\begin{figure}
%  \includegraphics[width=14cm]{figure6}\\
  \includegraphics[width=12cm]{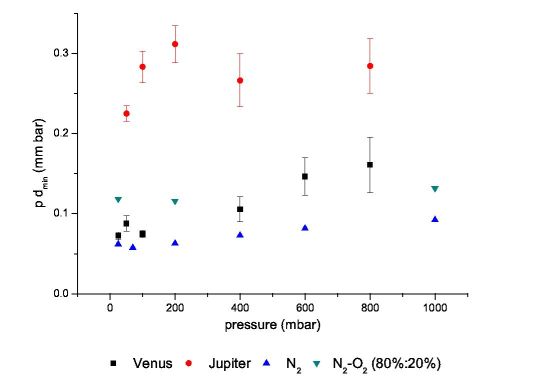}\\
  \caption{
     Reduced minimal streamer diameter, $p\cdot d_{min}$, as function of pressure, at room
     temperature.
     Red circles: experimental results in jovian atmosphere, the H$_2$-He mixture.
     Black squares: experimental results in venusian atmosphere, the CO$_2$-N$_2$ mixture.
     The errors are determined as the maximum between the image resolution and the standard error.
     Higher pressure measurements are dominated by the resolution error, while
     lower pressure measurements are dominated by the standard error.
     The pure nitrogen and air data (blue and green triangles respectively) are taken with permission from
     \cite{Nijdam10pure-gasses}.
  }
  \label{FIG: width}
\end{figure}

Figure~\ref{FIG: width} shows the \textit{reduced minimal diameter}, $p\cdot d_{min}$, as function
of pressure in the two gas mixtures. Here $p$ is the pressure and $d_{min}$ is the measured minimal
diameter. We can use pressure instead of density because the gas temperature is the same in all
experiments, namely room temperature. As expected, the reduced minimal diameter depends very weakly
on pressure. In the jovian mixture $p\cdot d_{min}\approx0.26\pm0.03$~mm$\cdot$bar and in the
venusian mixture $p\cdot d_{min}\approx0.09\pm0.03$~mm$\cdot$bar. These values are of the same
scale as values measured by \cite{Briels08sim} in air and nitrogen and by
\cite{Nijdam10pure-gasses} in argon and in oxygen--nitrogen mixtures of varying concentration. The
average reduced diameter reported by Nijdam \textit{et al.} in dry air is $\sim0.12$~mm$\cdot$bar.
Some of the values reported in that work are shown in Figure~\ref{FIG: width}.

We can summarize that the values of $p\cdot d_{min}$ of the streamer head in the CO$_2$-N$_2$
mixture, in air and in pure nitrogen are quite similar. In the H$_2$-He mixture the value of
$p\cdot d_{min}$ is twice as large. In section~\ref{subsec: overview} we discuss some possible
reasons for such a difference. It is likely that the jovian mixture is ionized more easily than the
mixtures that represent Venus and Earth, as well as pure nitrogen, creating wider and possibly
faster streamers.

We made preliminary measurements to estimate the minimal streamer velocity. We recall that
theoretically the streamer velocities do not depend on density when similarity laws apply, as
discussed by \cite{Briels08sim}. In all the pressures, in both mixtures, the measured minimal
velocity is of order of $10^5$~m/s. Our best estimation in the venusian mixture is
$\sim0.80\pm0.04\times10^5$~m/sec in 800 mbar, and in the jovian mixture it is
$1.00\pm0.05\times10^5$~m/sec in 50 mbar. \cite{Briels08sim} and \cite{Nijdam10pure-gasses} report
similar values in other gas mixtures, such as air. Velocities of $10^5$ to $10^7$ m/sec were
measured in sprite tendrils by means of high temporal resolution observations of terrestrial
sprites (\cite{Moudry02velocity, McHarg07fast, Stenbaek07velocity}).

Measurements by \cite{Briels08prmtrs} in air demonstrated that the velocity depends more strongly
on the applied voltage, than the diameter and above the inception voltage the velocity has a quite
large statistical error. We were not able to measure the minimal velocity in all the pressures,
however the lowest values we measured had the narrowest distribution. This is a clear indication
that the applied voltage was above the inception voltage.

According to \cite{Yair09}, sprites are expected on Venus at altitudes between 80 and~90~km above
the surface, and on Jupiter at $\sim$300~km above the 1~bar level. At these altitudes the pressure
is 5 to 0.4~mbar on Venus, and of the order of $10^{-3}$~mbar on Jupiter. The streamer minimal
diameter at such pressures according to our measurements are expected to be 0.2-0.02~m, and 300~m
respectively. The sprite tendrils may in fact be quite wider, as they do not need to be minimal.
For example, based on the value of $p\cdot d_{min}$ measured by \cite{Briels08sim} in air, minimal
sprite tendrils on Earth should be roughly 20~m wide. However observations report tendrils as wide
as~$\sim$150~m (\cite{Gerken03morphology}).

\section{Spectral Measurements}\label{sec: spectrum}

\subsection{Method}

We used two small spectrometers to determine the spectra emitted by the various discharges under
investigation. These spectrometers are sensitive in different wavelength regions: an Ocean Optics
HR2000 is sensitive between 177 and 622~nm and an Ocean Optics HR2000+ is sensitive between 420 and
820~nm. In the following, we refer to the spectra of these devices as \textit{UV-Vis} and
\textit{Vis} respectively. The corresponding sensitivity curves are shown in Figure~\ref{FIG:
sensitivity}. An optical fibre is used to get the light into the spectrometer. The acceptance angle
of this fibre is enough to capture light from the entire discharge region, when placed behind the
large quartz window (See Figure~\ref{fig:Setup-overview}). This includes the electrodes. We believe
that their radiation contribution to the spectrum is minor. In these discharges the electrodes
hardly heat up, and therefore do not emit planck radiation. The discharge around the electrodes
should not be that much different from the bulk. In most cases, the end of the optical fibre was
placed perpendicular to the large quartz window, in full view of the discharge region, but outside
the ITO window of the Faraday cage. However, in some cases, it was placed within the Faraday cage,
parallel to the large quartz window. In these latter cases, a mirror was used to direct the light
from the discharge to the fibre. This is needed to measure at wavelengths below 300~nm, which are
absorbed by the ITO window. The construction with the mirror was necessary because of the small
space between the ITO and the quartz windows. In order to get enough radiation from the discharge
to produce a spectrum with an acceptable signal to noise ratio, we needed to use long measurement
times, up to 100 minutes, and a high discharge repetition rate of 10~Hz. When the higher repetition
rate of~10~Hz is used, we do not use the equipment responsible for constantly renewing the gas in
the vessel, due to technical difficulties. It is possible that meta-stable molecules form in the
gas and influence the measured spectrum. The measurements shown here are averaged curves of 10
intervals of 60~sec integration time each.

Because of the very low intensity of the streamers in the two gasses, we had to use low pressures
and high voltages. The discharges we investigate look like those shown in the right column of
Figure~\ref{FIG: overview}. Most of the light emission in these discharges comes from the glow
phase, which is longer in duration and more intense. Nijdam~\textit{et al.}
(http://arxiv.org/abs/0912.0894v1, version 1) show that the spectra of streamers and of pulsed glow
in pure nitrogen are practically indistinguishable. We assume it is true for the planetary gasses
as well. Nonetheless, one should remember that the spectra discussed in this section are basically
those of a pulsed glow discharge that lasts a few micro-seconds.

Two spectra, using both spectrometers, were measured in each gas mixture.
The spectra presented in this paper are corrected for the sensitivity of the spectrometers. The
intensity scale is equal between the different spectra and spectrometers. Therefore, intensities
can be compared between different spectra and wavelengths. However, we have no absolute intensity
calibration. We normalise the spectra with the strongest lines in the venusian spectra, and with
the Balmer $\alpha$ line in the jovian spectrum. In the measured spectra we have excluded some
pixels that showed excessive noise levels in the calibration measurements. When more than two
neighbouring pixels are excluded, this section of the spectrum is removed from the spectrum. In
other cases the excluded pixels are represented by the average of their neighbours. In total about
75 points per spectrometer have been excluded.

\begin{figure}
%  \includegraphics[width=14cm]{figure7}\\
  \includegraphics[width=12cm]{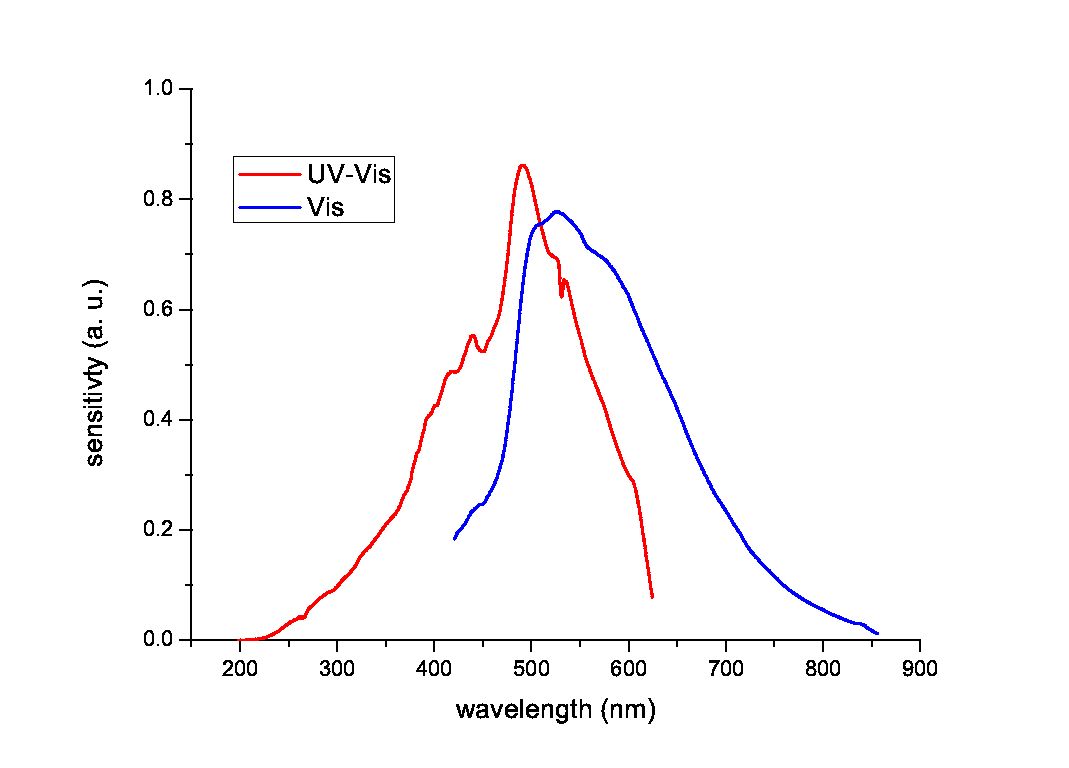}\\
  \caption{Normalized sensitivity curves of the two spectrometers.
  The spectrometer sensitivity curves have been acquired by means of calibrated deuterium
  and halogen lamps.
  }
  \label{FIG: sensitivity}
\end{figure}

In this section we report our observations on the streamer and glow discharge spectra in the two
planetary gasses, and the spark spectrum in the mixture that corresponds to Venus. For spectral
identification we use the tables of molecular spectra by \cite{Pearse1965}, and the NIST database
for atomic spectral lines (\cite{NIST-ASD}).

\subsection{Spectrum for sprites on Jupiter}

\begin{figure}
%  \includegraphics[width=14cm]{figure8}\\
  \includegraphics[width=12cm]{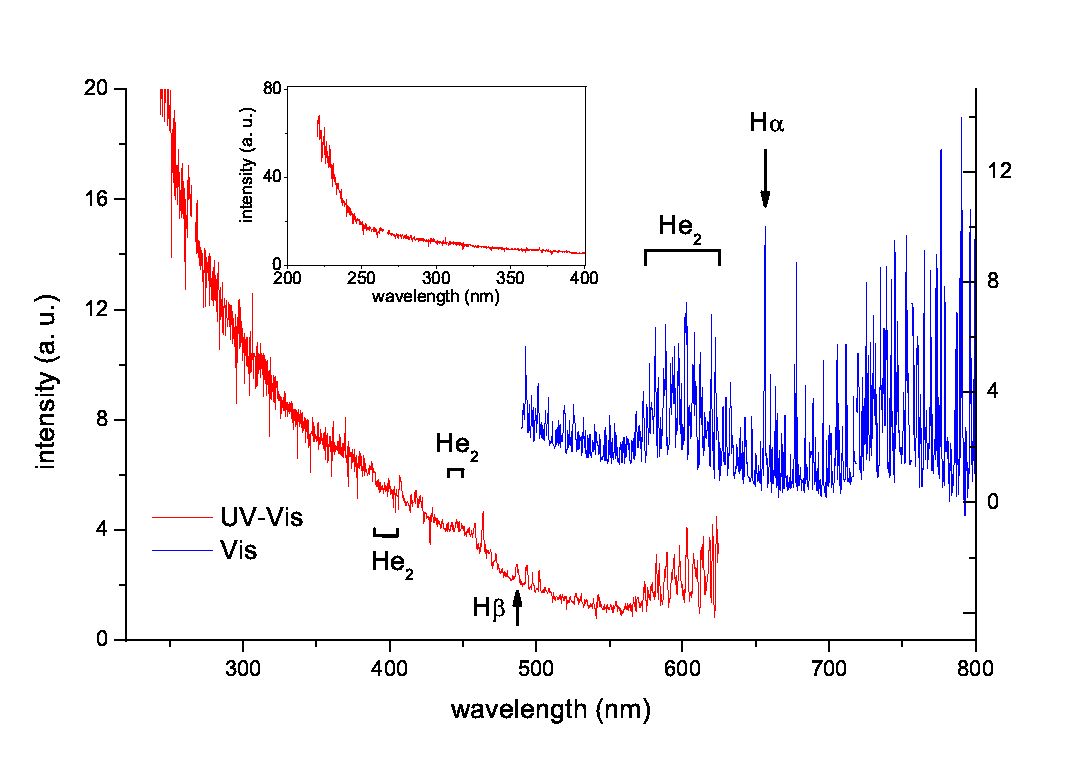}\\
  \caption{Spectrum of streamer and glow discharge in a gas mixture that simulates the jovian atmosphere,
     H$_2$-He - 89.8\%:10.2\%.
     The strong continuum is dominant in the {\it UV-Vis} range (shown in the inset).
     }
  \label{FIG: spectrum Jupiter}
\end{figure}

Figure~\ref{FIG: spectrum Jupiter} shows the spectrum obtained in the mixture that simulates
Jupiter's atmosphere, at a pressure of 50 mbar and a voltage peak of~$\sim$25 kV. These
measurements were taken with the optic fiber placed between the discharge vessel window and the ITO
glass, so radiation below 300 nm was recorded.

The dominant feature of this spectrum is the continuum in the {\it UV-Vis} range, with a higher
intensity at the lower wavelengths. It is similar in form to the UV continuum emission in H$_2$
reported by \citet{Lavrov99UV-cont}, and references therein.
Continuum in H$_2$-containing mixtures is created in transition from electronically-excited state
$a^3\Sigma^{+}_{g}$ of H$_2$ to the unstable state $b^3\Sigma^{+}_{u}$, which auto-dissociates
instantaneously with a photon emission, i.e. $a^3\Sigma^{+}_{g}\rightarrow
b^3\Sigma^{+}_{u}\rightarrow$ H + H + photon (Sergey Pancheshnyi in personal communication).
%
%It is associated with the $a^3\Sigma^{+}_{g}\rightarrow b^3\Sigma^{+}_{u}$ electronic transition in H$_2$.
%
In addition to this continuum, there are many spectral lines. We focus here on the most distinctive
features of the spectrum. The strong and narrow line at 656~nm, is the H$\alpha$ line. The second
line of the Balmer series, much less intense, is present as well. There are two regions of
particularly strong and dense lines, one at the wavelength range of 575-625~nm and another at
700-800~nm. The 575-625~nm band probably belongs to the He$_2$ molecular spectrum. The He$_2$
spectrum has two other such regions at wavelengths below 500~nm (marked in the Figure). These bands
are not apparent in our spectrum. The 700-800~nm band, we are currently unable to identify. A
possible candidate is the hydrogen molecule, which has a very complex and dense spectrum in the
visible range. However while many of the peaks of the H$_2$ spectrum lines coincide with peaks in
Figure~\ref{FIG: spectrum Jupiter}, many others seem to be absent or otherwise they are too weak
and difficult to identify, particularly in the 500-550nm range. A higher resolution measurement is
required to allow for un-ambiguous identification.

\subsection{Spectrum for sprites on Venus}

\begin{figure}
%  \includegraphics[width=14cm]{figure9}\\
  \includegraphics[width=12cm]{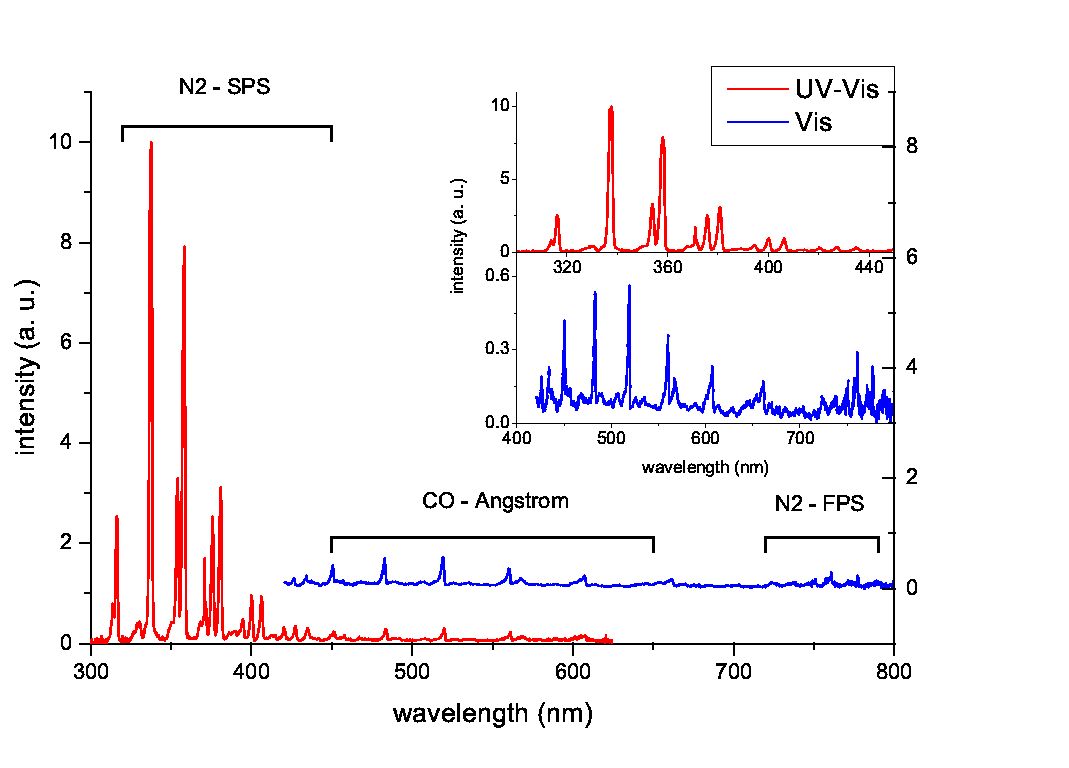}\\
  \caption{Spectrum of streamer and glow discharge in a gas mixture that simulates Venus's atmosphere,
     CO$_2$-N$_2$ -- 96.5\%:3.5\%.
     The main figure shows the entire spectrum to scale.
     The lines of N$_2$-SPS are shown in the top inset, in the UV-Vis range.
     The lines of the CO-{\AA}ngstrom system and N$_2$-FPS are shown in the bottom inset, scaled
     up.
     Wavelengths of most significant lines are indicated in the text.
     }
  \label{FIG: spectrum Venus}
\end{figure}

Figure~(\ref{FIG: spectrum Venus}) shows the spectrum obtained in the mixture that simulates the
venusian atmosphere, at a pressure of 50 mbar and applied voltage of~$\sim$43~kV. The optic fiber
is placed behind the ITO window, so no spectral lines can be observed below the wavelength of 300
nm.

The most dominant feature of this spectrum is the N$_2$ second positive band of triple heads in the
range 300 to 450 nm (N$_2$-SPS). The strongest lines of this system are found at 316, 337, 354,
358, 376 and 381~nm. All the lines have widths of $\sim$2~nm. The CO {\AA}ngstrom system is clearly
visible in the {\it Vis} spectrum, with lines at 451~nm, 483~nm, 519~nm, 561~nm, 607~nm and 662~nm,
all degraded to the violet. This system is considerably weaker that the N$_2$-SPS band. The line at
567 nm is possibly one of the triplets of the CO triplet system. Other triplets of this system
might be present as well, but many of them are very close to CO {\AA}ngstrom lines, making it
difficult to identify them (such as the wide line at 600-607~nm), others are too weak to be
un-equivocally identified. Several very weak heads of the N$_2$ first positive band (N$_2$-FPS),
are found at the upper edge of the spectrum, at the wavelength range 725-790~nm.

It is somewhat surprising that we observe primarily nitrogen lines in this spectrum. Previous work
by Goto and Ohba (unpublished report 2008) in hot plasma, found a significant CO$_2$ signature in
the spectrum, while ~\cite{Borucki96spec} found mainly atomic oxygen and carbon lines. However,
according to the literature (\cite{Pearse1965}) even a very small amount of nitrogen mixed into a
gas can produces strong N$_2$ lines in some circumstances, and these nitrogen lines can be
considerably stronger than any other feature of the spectrum. This seems to be the case with the
streamer spectrum in the venusian atmosphere examined here. This is an indication that pure CO$_2$
may not the best choice when one wishes to simulate such discharges in the venusian atmosphere, and
nitrogen must be taken into account. The hot plasma spectrum is discussed in more detail in the
following section.

\subsection{Spectrum for lightning on Venus}\label{sec: spark}

By a simple modification, the system described in section~\ref{subsec: set-up} can create sparks in
the discharge gap. The resistor R$_3$ in Figure~\ref{fig:Circuit-simple}, through which most of the
current during a voltage pulse, is replaced by a resistor with a higher resistance (a~1~k$\Omega$
resistor is replaced by a~6~M$\Omega$ resistor). The original resistor is chosen in such away that
most of the current flows through the R$_3$ branch of the circuit, rather than through the
discharge gap, therefore the high voltage on the discharge gap falls very rapidly, within several
micro-seconds. With the stronger resistor in place, the high voltage on the electrodes persists for
a time long enough for a spark discharge to occur, if the gas in the vessel is dense enough. The
spark discharge is akin to lightning on a small scale.

\begin{figure}
  \includegraphics[width=12cm]{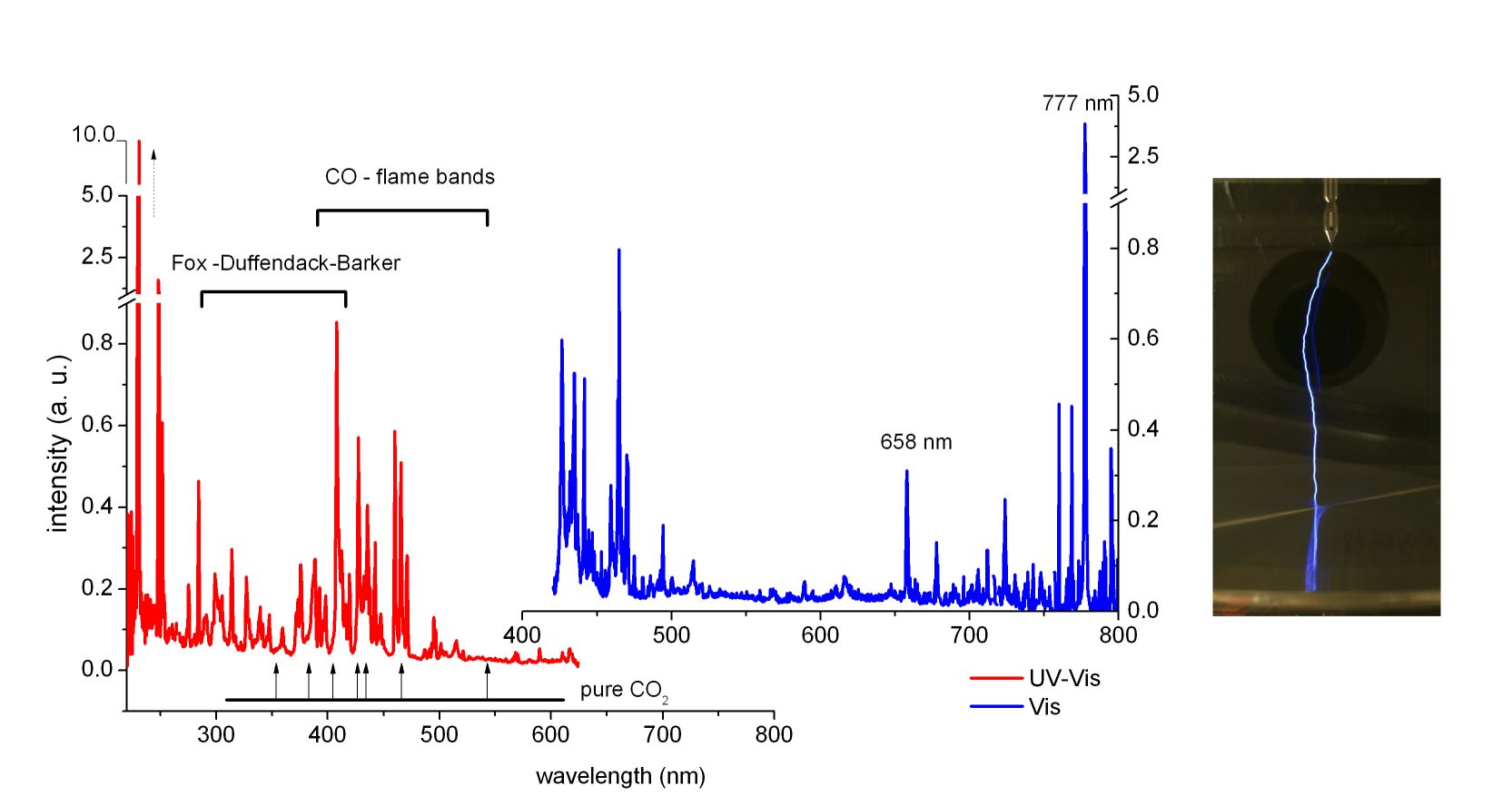}
  \caption{Left: Spectrum of a spark discharge in a gas mixture that simulates the venusian atmosphere,
     CO$_2$-N$_2$ -- 96.5\%:3.5\%.
     Arrows indicate the CO$_2$ spectral lines observed by Goto and Ohba (unpublished report 2008).
     A strong OI line is
     at 777~nm, two CO$^+$-FNS lines are at 230~nm and 250~nm.
     Right: a color photograph of the spark, taken with a SONY $\alpha$300 camera that was placed
     in a separate Faraday cage for protection reasons.
     The bright channel is the spark.
     The faint ``channel'', and the smudged blue light at the bottom of the picture are reflections from
     the plate electrode, and the three windows between the spark and the camera lens (the gas vessel
     window and two ITO windows in both faraday cages).
     }
  \label{FIG: spectrum Spark}
\end{figure}

We used this setting to create sparks in the mixture that represents Venus's atmosphere,
CO$_2$-N$_2$, and recorded its spectrum using the equipment and methods described above. The optic
fiber was placed between the vessel quartz window and the ITO window, so wavelengths shorter than
300~nm were observed. The pressure in the vessel was 200~mbar, and a voltage of~$\sim$50~kV was
used. The sparks created in this way are significantly brighter than the cold plasma discharges
discussed above. As a result we could measure the spectrum with considerable shorter exposure
times; 10 seconds for the {\it UV-Vis} range, and 30 seconds for the {\it Vis} range were used. The
curves shown in Figure~\ref{FIG: spectrum Spark}, are the averaged result of 10 measurements.

When one compares the spectra in figures~\ref{FIG: spectrum Venus} and~\ref{FIG: spectrum Spark},
the difference is apparent.  The strongest lines of the cold plasma spectrum are concentrated in
the wavelength range 300-400~nm, and these are nitrogen lines. There are no CO$_2$ lines, and the
CO lines are very weak. The spark spectrum, on the other hand, has a concentration of lines between
400 and 500~nm, and several strong and narrow lines; the atomic oxygen line at $\sim$777~nm, and
two lines in the UV, $\sim$230~nm and $\sim$250~nm, that probably belong to the CO$^+$ first
negative system. Many of the other lines in this spectrum correspond with the carbon monoxide flame
bands ($^1B_2-X^1\Sigma^+$), which are in fact CO$_2$ emissions, and also with the Fox-Duffendack
and Barker's system ($A^2\Pi-X^2\Pi$), which are CO$_2^+$ emissions.

Goto and Ohba (unpublished report 2008) performed  measurements of the hot plasma spectrum in pure
CO$_2$ at several pressures. Their results are similar to the spark spectrum shown here. They
identified several wide CO$_2$ lines. The spectrum in Figure~\ref{FIG: spectrum Spark} has a higher
resolution, and therefore many more lines are observed. These lines are grouped into several bands
that agree well with the wide lines of the 100~Torr (133~mbar) spectrum measured by Goto and Ohba,
and which are indicated with arrows in the figure. These bands in our spectrum are located roughly
at 375~nm, 390~nm, 415~nm, 440~nm and 470~nm, and all have widths between 10 and 20~nm. We also
observe a strong oxygen line at~777~nm which was also reported by Goto and Ohba as well as by
\cite{Borucki96spec}.

It is interesting to note in this respect, that 6 of the 7 flashes reported by \cite{Hansell95} in
their ground observation of Venus's night-side were found using a 777.4~nm filter and one more
flash was observed using a 656.3~nm filter. Our spectrum has a strong OI line at~$\sim$777~nm, and
a clear and narrow line at 658~nm.

\section{Conclusions}

We have observed streamers in a previously unexplored set of gasses in controlled laboratory
settings. These gasses simulate the atmospheres of Venus and Jupiter-like planets. We have
demonstrated that streamer discharges are possible in these gasses, which gives firmer ground to
our hope of observing sprites on these planets. We have explored some features of these discharges,
such as the inception voltages, the minimal diameter and velocity of the streamer heads and their
intensity as compared to the optical brightness. of streamers in air. We find that the streamers in
these new gasses follow scaling laws as expected. We demonstrate that the reduced minimal diameter,
$p\times d_{min}$ does not depend on pressure. We also find that the streamers in the different
gasses are rather similar in their appearance, albeit differences in branching, intensity and the
propagation path (straight in air and the jovian mixture but a little jagged in the venusian
mixture). The reduced diameter in all the gasses is of the same order of magnitude. In the venusian
mixture it is very close to the reduced diameter in air that was reported by \cite{Briels08sim} and
by \cite{Nijdam10pure-gasses}, but it is twice as wide in Jupiter. Hence, the diameter is
influenced by the composition of the gas. Some considerations on the physical origin of these
differences, like electron friction and photoionization versus background ionization, are discussed
in section~\ref{subsec: overview}. Based on our reduced minimal diameter measurements, and the
sprite altitudes estimated by~\cite{Yair09}, we predict that sprite tendrils on Jupiter will be at
least 300~m wide, and of the order of a meter wide on Venus.

Most important in terms of the search for planetary sprites, is the fact that the streamers in both
mixtures emit much less light than in air, and that they require higher inception voltages,
particularly in the venusian mixture. Therefore, planetary sprites can be expected to have similar
morphology to terrestrial sprites, but they might be significantly weaker in optical brightness and
require a larger charge moment of the parent lightning. These considerations should be taken into
account if a more detailed scheme for predicting sprite altitudes than the one proposed by
\cite{Yair09}, is to be employed.

To maximize the chances of discovering planetary sprites by optical observations, one should focus
on the specific spectral lines expected in these discharges. We have examined the optical spectra
of streamers and glow in the two atmospheres, in the visual and the near UV range. In the venusian
atmosphere we find strong N$_2$ lines from the second positive band, as well as considerably weaker
CO lines. We find that this spectrum is significantly different from the hot plasma spectrum found
by \cite{Borucki96spec} in a similar gas mixture, being populated by many more spectral lines. It
is also significantly different from the spectrum measured by Goto and Ohba (unpublished report
2008) in pure CO$_2$, and from our own measurements of the spark spectrum in the N$_2$-CO$_2$
mixture. We find that in the case of a cold plasma discharge the presence of nitrogen in the
atmospheric gas is important and should not be neglected. In case of a hot plasma discharge, a
spark, CO$_2$ and CO lines are dominant. Among the strongest lines in our spark spectrum, it is
worth mentioning the lines at 777~nm and 658~nm. These lines are close to the wavelengths proposed
by \cite{Hansell95} for lightning observations on Venus. We observe also two very strong lines at
230~nm and 250~nm, which may be useful as well. However, these two lines are at the edge of our
spectrum, where the apparatus sensitivity is low. It is worthwhile to look more closely at this
wavelength.

In Jupiter's atmosphere the spectrum is more complex, consisting both of continuum radiation and
very dense spectral lines.
The Balmer alpha line, is a common feature of spectra of gasses that contain hydrogen, and can not
be considered a characteristic of the cold plasma process spectrum. For example,
\cite{Borucki96spec} find this line as the strongest feature of the hot plasma spectrum in a
similar gas mixture.
On the other hand, we see a dominant continuum spectrum, which is not found in Borucki's work.
Moreover, there are several regions of dense band structure, which are characteristic both of
Helium molecular and atomic spectra, and of the hydrogen molecular spectrum. Calibrated
measurements of the jovian spectrum, may yield information on the electron density and energies in
these discharges. Further investigation of the jovian and dry-air spectra is under investigation
and some of the results are about to be published by Nijdam \textit{et al.}
(http://arxiv.org/abs/0912.0894v1, version 1).

%\begin{acknowledgments}

This research is supported by the Israeli Science Foundation grant 117/09. Support for experiments
conducted by the first author was received from COST Action P-18, ``Physics of Lightning Flash and
its Effects'' as part of a Short Term Scientific Mission (STSM). SN acknowledges support by
STW-project 06501, part of the Netherlands' Organization for Scientific Research NWO.

%\end{acknowledgments}

%\bibliographystyle{agufull08}
\bibliographystyle{plainnat} %Harvar style
%\bibliography{mybib}

%\end{article}

\end{document}